\documentstyle[12pt,epsfig]{article}

\newcommand{\lsim}{\raisebox{-0.13cm}{~\shortstack{$<$ \\[-0.07cm] $\sim$}}~}
\newcommand{\gsim}{\raisebox{-0.13cm}{~\shortstack{$>$ \\[-0.07cm] $\sim$}}~}

\newcommand{\ra}{\rightarrow}

\newcommand{\s}{\\ \vspace*{-3.5mm} }

\newcommand{\non}{\nonumber}
\newcommand{\beq}{\begin{eqnarray}}
\newcommand{\eeq}{\end{eqnarray}}

\newcommand{\tb}{\mbox{tg}\beta}

\begin{document}

\begin{flushright}
PM/98--06 \\
GDR--S--06
\end{flushright}

\vspace{1cm}

\begin{center}

{\large\sc Squark Effects on Higgs Boson Production}

\vspace*{4mm} 

{\large\sc and Decay at the LHC}

\vspace*{9mm}

{\sc Abdelhak DJOUADI} 

\vspace*{9mm}

Laboratoire de Physique Math\'ematique et Th\'eorique, UMR--CNRS, \\
Universit\'e Montpellier II, F--34095 Montpellier Cedex 5, France. \\
E-mail: djouadi@lpm.univ-montp2.fr

\end{center}

\vspace*{2cm}

\begin{abstract}
\normalsize

\noindent
In the context of the Minimal Supersymmetric extension of the Standard Model,
I discuss the effects of relatively light top and bottom scalar quarks 
on the main production mechanism of the lightest SUSY 
neutral Higgs boson $h$ at the LHC, the gluon--gluon fusion mechanism $gg \ra 
h$, and on the most promising discovery channel, the two--photon decay mode 
$h \ra \gamma \gamma$. In some areas of the parameter space, the top and
bottom squark contributions can strongly reduce the production cross section 
times the branching ratio. 

\end{abstract}

\newpage

\subsection*{1. Introduction}

In the Minimal Supersymmetric extension of the Standard Model (MSSM) \cite{R1},
the electroweak symmetry is broken with two Higgs--doublet fields, 
leading to the existence of five physical states \cite{R2}: two CP--even Higgs 
bosons $h$ and $H$, a CP--odd Higgs boson $A$ and two charged Higgs particles 
$H^\pm$. In the theoretically well motivated models, such as Supergravity 
models, the MSSM Higgs sector is in the so called decoupling regime \cite{R4} 
for most of the SUSY parameter space allowed by present data 
constraints \cite{R5}: the heavy CP--even, the CP--odd  and the charged Higgs
bosons are rather heavy and almost degenerate in mass, while the lightest 
neutral CP--even Higgs  particle reaches its maximal allowed mass value 
$ M_h \lsim $ 80--130 GeV \cite{mh} depending on the SUSY parameters. 
In this scenario, the $h$ boson has almost the same properties as the SM 
Higgs boson [and a lower bound on its mass, $M_h \gsim 88$ GeV, is 
set \cite{R5} by the negative LEPII searches] and would be the sole Higgs 
particle accessible at the LHC. \s

At the CERN Large Hadron Collider (LHC), the most promising channel \cite{LHC} 
for detecting the Standard Model (SM) Higgs boson $H^0$ in the mass range
below $\lsim 150$ GeV, is the rare decay into two photons \cite{R7a} 
$H^0 \rightarrow \gamma \gamma$, with the Higgs particle dominantly 
produced via the top quark loop mediated gluon--gluon fusion mechanism 
\cite{R7b,R7c}  $gg  \rightarrow H^0$. [Two other channels can also be used in 
this mass  range \cite{LHC}: the production in association with a $W$ boson, or
with top quark pairs with $t \ra bW$; although the cross sections are 
smaller  compared to the $gg \ra H^0$ case, the backgrounds are also  small if 
one requires a lepton from the decaying $W$ bosons as an additional tag, 
leading to a cleaner signal.] The two LHC collaborations expect to detect 
the narrow $\gamma \gamma$ peak in the intermediate Higgs mass range, 80 GeV
$\lsim M_{H^0} \lsim 150$ GeV for the CMS collaboration and 100 GeV $\lsim 
M_{H^0} \lsim 140$ GeV for the ATLAS collaboration, with an integrated 
luminosity $\int {\cal L} \sim 100$ fb$^{-1}$ corresponding to one year LHC 
high--luminosity running \cite{R6}. \s

In the Standard Model, the Higgs--gluon--gluon vertex is mediated by heavy 
quark [mainly top and to a lesser extent bottom quark] loops, while the rare 
decay into two--photons is mediated by $W$--boson and heavy fermion loops, 
with the $W$--boson contribution being largely 
dominating. In the MSSM however, additional contributions are provided by 
SUSY particles: squark loops in the case of the $hgg$ vertex, and charged 
Higgs boson, sfermion and chargino loops in the case of the $h \ra \gamma 
\gamma$ decay. In the latter case \cite{gamma}, the contributions of 
$H^\pm$ bosons, sleptons and the scalar partners of the light quarks, 
and to a lesser extent charginos, are small given the experimental bounds 
on the masses of these particles \cite{R5} [the contribution of chargino 
loops can exceed the 10\% level for masses close to 100 GeV, but becomes 
smaller with higher masses]. Only the contributions of relatively light 
scalar top quarks, and to a lesser extent bottom squarks, can alter 
significantly the loop induced $hgg$ and $h \gamma \gamma$ vertices.\s 

In this note, I discuss the effects of the scalar $\tilde{t}$ and $\tilde{b}$
quark loops on the cross section for the production process $gg \ra h$ 
at the LHC and the decay mode $h \ra \gamma \gamma$. I will mainly focus 
on the case of the $h$ boson in the decoupling regime. I will also briefly 
discuss the case of the heavy CP--even and CP--odd Higgs production 
in the gluon--fusion mechanism. 

\subsection*{2. Physical Set--Up} 

As mentioned previously, $\tilde{t}$ and $\tilde{b}$ loops can affect 
significantly the $hgg$ and $h\gamma \gamma$ vertices, and the reason is 
twofold: the lightest $\tilde{t}$ and $\tilde{b}$ squarks can be relatively 
light, and their couplings to the $h$ boson strongly enhanced. \s

The current eigenstates, $\tilde{q}_L$ and $\tilde{q}_R$, mix to give 
the mass eigenstates $\tilde{q}_1$ and $\tilde{q}_2$  which are 
obtained by diagonalizing the following mass matrices
\begin{equation}
{\cal M}^2_{\tilde{q}} = \left(
  \begin{array}{cc} m_{\tilde{q}_L}^2 + m_q^2 + D_L^q & m_q \, \tilde{A}_q \\
                    m_q \tilde{A}_q & m_{\tilde{q}_R}^2 + m_q^2+ D_R^q 
\end{array} \right)
\end{equation}
where the off--diagonal entries are $\tilde{A}_t= A_t-\mu/\tb$ and $\tilde{A}_b
= A_b - \mu\tb$, with  $\tb$ the ratio of the vacuum expectation values of the 
two--Higgs fields  which break the electroweak symmetry, and $A_q$ and $\mu$ 
the soft--SUSY  breaking trilinear squark coupling and Higgs mass parameter, 
respectively. $m_{\tilde{q}_L}$ and $m_{\tilde{q}_R}$ are the left-- and 
right--handed soft--SUSY breaking scalar quark masses  which, in models with 
universal scalar masses at the GUT scale, are approximately equal to the 
common squark mass $m_{\tilde{q}}$. The $D$ terms, in units of $M_Z^2 \cos 
2\beta$ are given in terms of the  electric charge and the weak isospin of 
the squark by: $D_L^q= I^{3}_q -e_q \sin^2\theta_W$ and $D_R^q= e_q\sin^2
\theta_W$. \s

In the case of the top squark, the mixing angle $\theta_{\tilde{t}}$
is proportional to $m_t \tilde{A}_t$ and can be very large, leading to a scalar
top quark $\tilde{t}_1$ much  lighter than the $t$--quark and all other scalar 
quarks. In this case, the lightest top squark will not decouple from the 
$h\gamma \gamma$ and $hgg$ amplitudes. For large values of $\tb \sim m_t/m_b$, 
the mixing in the $\tilde{b}$ sector can also be important, leading to a 
relatively light $\tilde{b}_1$ squark. The experimental limits on the squark 
masses from negative LEPII and Tevatron searches are \cite{R5}: $m_{\tilde{t}_1
}, m_{\tilde{b}_1} \gsim 75$ when squark mixing is included [the bound on 
$m_{\tilde{b}_1}$ from the Tevatron does not hold in the case of large mixing] 
and for the other approximately degenerate squarks, $m_{\tilde{q}} \gsim 230$ 
GeV. \s

Normalized to $2M_Z^2(\sqrt{2}G_F)^{1/2}$, the couplings of top and bottom 
squark pairs to the $h$ boson read in the decoupling regime, 
\begin{eqnarray}
g_{h \tilde{q}_1 \tilde{q}_1 } &=& - \cos 2\beta \left[ 
I_q^3 \cos^2 \theta_{\tilde{q}} - e_q \sin^2 \theta_W \cos 2
\theta_{\tilde{q}} \right] 
- \frac{m_q^2}{M_Z^2} + \frac{1}{2} \sin 2\theta_{\tilde{q}} 
\frac{m_q \tilde{A}_q } {M_Z^2} \nonumber \\
g_{h \tilde{q}_2 \tilde{q}_2 } &=& - \cos 2\beta \left[ 
I_q^3 \sin^2 \theta_{\tilde{q}} - e_q \sin^2 \theta_W \cos 2
\theta_{\tilde{q}} \right] 
- \frac{m_q^2}{M_Z^2} - \frac{1}{2} \sin 2\theta_{\tilde{q}} 
\frac{m_q \tilde{A}_q } {M_Z^2}
\end{eqnarray}
and involve components which are proportional to $\tilde{A}_q$. In the case
of stop squarks, for large values of the parameter $\tilde{A}_t$ which 
incidentally make the $\tilde{t}$ mixing angle  maximal, $|\sin 2 \theta_{
\tilde{t}}| \simeq 1$, the latter terms can strongly  enhance the $g_{h
\tilde{t}_1 \tilde{t}_1 }$ coupling and make it larger than the  top quark 
coupling of the $h$ boson, $g_{htt} \propto m_t/M_Z$. This component and the
$m_t^2/M_Z^2$ component of the coupling would result in a contribution to 
the $h gg$ and $h \gamma \gamma$ vertices that is comparable or even larger 
than the top quark contribution. Here again,  the $h\tilde{b}_1 \tilde{b}_1$ 
couplings can also be very strongly enhanced for large $\tb$ values, and could 
alter significantly the $hgg$ and $h\gamma \gamma$ vertices. \s

In this note, both the low and large $\tb$ cases will be discussed, and for 
illustration the values $\tb \sim 2.5$ and $\tb \sim  50$ will be 
used\footnote{In the case of
low [$\tb \sim 2$] and large [$\tb \sim m_t/m_b$] values which are favored 
by Yukawa coupling unification, assuming the decoupling limit for the $h$ 
boson is further justified: if the $h$ boson is not discovered at LEPII at 
$\sqrt{s} \sim 200$ GeV, values of $\tb \lsim 2$ will be ruled out and the 
$h$ boson is SM--like for allowed $\tb$ values close to this limit; for large 
$\tb$, Tevatron data imply \cite{man} $M_H \sim M_A \gsim 150$ GeV, and the 
$h$ boson should again be SM--like.}.
However, the analysis applies for any $\tb$ if, as it will be the case,
$\tilde{A}_q$ is used as the input parameter [since the $\tb$ dependence is 
hidden in $\tilde{A}_q$]. The only difference, when using different $\tb$ 
values, would be the different value of the lightest $h$ boson mass that
is obtained in the decoupling limit. \s

The expression of the partial width for the decay $h \ra gg$, including
only the contributions of the top/bottom quarks and their spin-zero 
partners, is given by
\beq
\Gamma (h \ra gg) = \frac{G_F \alpha^2_s 
M_{h}^3}{64  \sqrt{2} \pi^3 } \left| \, 
\sum_Q A_Q (\tau_Q)  +  \sum_{\tilde{Q}} g_{h \tilde{Q} \tilde{Q}} \, 
\frac{M_Z^2} {m_{\tilde{Q}}^2} \, A_{\tilde{Q}} (\tau_{\tilde{Q}}) \, 
\right|^2
\eeq
where the scaling variable $\tau_i$ is defined as $\tau_i = M_{h}^2/
4m_i^2$ with $m_i$ the mass of the loop particle, and the amplitudes 
$A_i$ are  
\beq
A_{Q}(\tau) &=& - 2 [\tau+(\tau-1)f(\tau)]/\tau^2 \non \\
A_{\tilde{Q}} (\tau) &=& [\tau -f(\tau)]/\tau^2 
\eeq
with the function $f(\tau)$ defined by
\begin{equation}
f(\tau) = \left\{ \begin{array}{ll} 
{\rm arcsin}^2 \sqrt{\tau} & \tau \leq 1 \\
-\frac{1}{4} \left[ \log \frac{1 + \sqrt{1-\tau^{-1} } }
{1 - \sqrt{1-\tau^{-1}} } - i \pi \right]^2 \ \ \ & \tau >1 
\end{array} \right. 
\end{equation} 
In the SM,  the main contribution comes from the top quark for which one can 
take the limit $A_Q \ra -4/3$. In the case of squarks, only $\tilde{t}$
and $\tilde{b}$ contribute, and below the particle threshold $M_h <2 
m_{\tilde{Q}}$, the amplitudes $A_{\tilde{Q}}$ are real and reach the value
$A_{\tilde{Q}} \ra -1/3$ for heavy loop masses. The sum of the contributions
of the scalar partners of the first and second generation quarks is zero. \s

The cross section for $h$ production in the $gg$--fusion mechanism $\sigma(gg 
\ra h$) is directly proportional to the gluonic decay width $\Gamma(h \ra gg)$.
The latter cross section is affected by large QCD radiative corrections 
\cite{R7c}; however the corrections are practically the same for quark and 
squark loops, and since only deviations compared to the SM case will be 
considered here, they drop out in the ratios. The partial width for the decay
$h \ra \gamma \gamma$ can be found e.g. in Ref.~\cite{gamma}. The QCD 
corrections are small in the case of the $h \ra \gamma \gamma$ decay and 
can be neglected. The $\gamma \gamma$ and $gg$ decay widths of the $h$ boson 
are evaluated numerically with the help of an adapted version of the program 
HDECAY \cite{hdecay}.

\subsection*{3. Numerical Results} 

Figs.~1--3 show the deviations from their SM values of the partial 
decay widths of the $h$ boson into two photons and two gluons as well as 
their product which gives the cross section times branching ratio 
$\sigma(gg \ra h \ra \gamma \gamma)$. The quantities $R$ are defined as 
the partial widths including the SUSY loop contributions [all charged SUSY 
particles for $h \ra \gamma \gamma$ and squark loops for $h \ra gg$] 
normalized to the partial decay widths without the SUSY contributions, 
which in the decoupling limit correspond to the SM contributions: 
$R=\Gamma_{\rm MSSM}/\Gamma_{\rm SM}$. \s

Since, as discussed previously, the main SUSY contribution for small values
of $\tb$ are due to $\tilde{t}$ loops, the loop contributions are shown in  
Figs.~1--2 as a function of $\tilde{A}_t$ for $\tb=2.5$ and the values 
$m_{\tilde{t}_1}=200$ GeV (Fig.~1) and  $m_{\tilde{t}_1}=165,400$ and 600 GeV 
(Fig.~2) for the $\tilde{t}_1$ mass [which then fixes the parameters
$m_{\tilde{t}_L} \simeq m_{\tilde{t}_R} \simeq m_{\tilde{q}}$]. The other 
parameters are chosen as $M_2=-\mu=250$ and 500 GeV for the scenarii 
$m_{\tilde{t}_1} \leq 200$ GeV and $>200$ GeV respectively; the choice of 
these two different values is motivated by the requirement that the lightest 
neutralino must be lighter than $\tilde{t}_1$ in each scenario. \s

Concentrating first on the case $m_{\tilde{t}_1}=200$ 
GeV, for small values of $\tilde{A}_t$ there is no mixing in the stop sector 
and the dominant component of the $h\tilde{t} \tilde{t}$ couplings,
eq.~(2), is the one proportional to $m_t^2/M_Z^2$ [here, both $\tilde{t}_1$ 
and $\tilde{t}_2$ contribute since their masses
and couplings to $h$ are almost the same]. The sign of this component, 
compared to the $ht\bar{t}$ coupling, is such that the top and 
stop contributions interfere constructively in the $hgg$ and $h\gamma \gamma$
amplitudes. This leads to an enhancement of the $h \ra gg$ decay width
up to $60\%$ in the MSSM. However, the $h \ra \gamma \gamma$ decay width is 
dominated by the $W$ amplitude which interferes destructively with the top and
stop quark amplitudes [there is also a small contribution  
from chargino loops in this scenario since the $\chi_1^+$ mass is $\simeq 
230$ GeV] and the $\tilde{t}$ contributions reduce the $h \ra \gamma 
\gamma$ decay width by an amount up to $-20\%$. The product R($gg \ra \gamma 
\gamma$) in the MSSM is then enhanced by a factor $\sim 1.2$ in this case. \s

With increasing $\tilde{A}_{t}$, the two components of the $h\tilde{t}_1 
\tilde{t}_1$ coupling [which have opposite sign because $\sin2\theta_{\tilde{t}}
\propto m_t \tilde{A}_t$ in eq.~(2)] interfere 
destructively and partly cancel each other, resulting in a rather small 
stop contribution. For a value  $\tilde{A}_t \sim 400$ GeV, $g_{h\tilde{t}_1 
\tilde{t}_1} \sim 0$ and the $\tilde{t}_1$ contributions to the 
$hgg$ and $h\gamma \gamma$ amplitudes vanish [here, $\tilde{t}_2$ is too heavy 
to contribute]. 
For larger values of $\tilde{A}_{t}$, the second component of the $h\tilde{t}_1 
\tilde{t}_1$ coupling becomes the most important one, and the $\tilde{t}_1$
loop contribution [$\tilde{t}_2$ is too heavy to contribute] 
interferes destructively with the one of the top quark.
This leads to an enhancement of R$(h \ra \gamma \gamma)$ and a reduction of
R$(gg \ra h)$. However, the reduction of the latter is much stronger than
the enhancement of the former [recall that the $W$ contribution in the
$h \ra \gamma \gamma $ decay is much larger than the top contribution]
and the product R($gg \ra \gamma \gamma$) decreases with increasing
$\tilde{A}_t$. For $\tilde{A}_t$ values of about 1.5 TeV, the signal 
for $ gg\ra h \ra \gamma \gamma$ in the MSSM is smaller by a factor of 
$\sim 5$ compared to the SM case\footnote{Note that despite of the large
splitting between the two stops and the sbottom that is generated by 
large values of $\tilde{A}_t$, the contributions of the $(\tilde{t},
\tilde{b}$) isodoublet to high--precision observables stay below the 
acceptable level. For instance, even for $\tilde{A}_t\sim 1.7$ TeV,
the contribution to the $\rho$ parameter is smaller than $3.10^{-3}$ 
which approximately corresponds to a 2$\sigma$ deviation from the SM 
expectation \cite{drho}.}. \s

Fig.~2 shows the deviation  R$(gg \ra \gamma \gamma)$ with the same parameters
as in Fig.~1 but with different $\tilde{t}_1$ masses, $m_{\tilde{t}_1}=
165,400$ and 600 GeV, and to ensure an LSP lighter than $\tilde{t}_1$, 
with $M_2=-\mu=500$ GeV for $m_{\tilde{t}_1}\geq 400$ GeV.   
For larger masses, the top squark contribution $\propto 1/ m_{\tilde{t}_1}^2$, 
will be smaller than in the previous case. In the no--mixing case, the 
enhancement (reduction) of the $hgg (h\gamma \gamma)$ 
amplitude is only of the order of 10\% for $m_{\tilde{t}_1}\simeq 400$ GeV,
and leads to an almost constant cross section times branching ratio for the 
$gg \ra h \ra 
\gamma \gamma$ process compared to the SM case. Again the stop contribution 
vanishes for some intermediate value of $\tilde{A}_t$, and then 
increases again in absolute value for larger $\tilde{A}_t$. However, 
for $m_{\tilde{t}_1}\simeq 400$ GeV,
the effect is less striking compared to the case of $m_{\tilde{t}_1}=200$ GeV, 
since here $\sigma(gg \ra h) \times {\rm BR}(h \ra \gamma \gamma$) drops by 
less than a factor of 2, even for extreme values of $\tilde{A}_t \sim 2.5$ TeV.
As expected, the effect of the top squark loops will become less important if 
the $\tilde{t}_1$ mass is increased further to 600 GeV for instance. 
In contrast, if the stop mass is reduced to $m_{\tilde{t}_1}\simeq 165$ GeV,
the drop in  R$(gg \ra \gamma \gamma)$ will be even more important:
for  $\tilde{A}_t \sim 1.5$ TeV, the $gg \ra \gamma \gamma$ cross section
times branching ratio including stop loops is an order of magnitude smaller 
than in the SM. For $\tilde{A}_t \sim 1.3$ TeV, the stop amplitude almost 
cancels completely the top and bottom quark amplitudes; the non--zero value
of R$(gg \ra \gamma \gamma)$ is then due to the imaginary part of the bottom 
quark contribution.  \s

Note that $M_h$ varies with $\tilde{A}_t$, and no constraint on $M_h$ has been 
set in Figs.~1--2. Requiring $M_h \gsim 90$ GeV, the lower range $\tilde{A}_t 
\lsim 350$ GeV and the upper ranges $\tilde{A}_t \gsim 1.5 (2.3)$ TeV for 
$m_{\tilde{t}_1}=200 (400)$ GeV for instance, are ruled out. [This is due 
to the fact that the maximal value of $M_h$  for a given $\tb$ 
and a common scalar mass $m_{\tilde{q}}$, which here is fixed in terms of 
$m_{\tilde{t}_1}, \tb$ and $\tilde{A}_t$, the $h$ boson mass increases with 
increasing $\tilde{A}_t$ and reaches a maximal value for $\tilde{A}_t \simeq 
\sqrt{6} m_{\tilde{q}}$; when $\tilde{A}_t$ exceeds this value, the maximal 
value of the $h$ boson mass will start decreasing.] This means that the 
scenario where R$(gg \ra \gamma \gamma) >1$, which occurs only for small 
values of $\tilde{A}_t \lsim 300$ GeV for $m_{\tilde{t}_1}=200$ GeV
is ruled out for $M_h \gsim 90$ GeV. Therefore, if this constraint is 
implemented, 
the cross section times branching ratio for the $gg \ra \gamma \gamma$ 
process in the MSSM will always be smaller than in the SM case, making
more delicate the search for the $h$ boson at the LHC with this 
process\footnote{Note that when these contributions are significant, the 
process \cite{hstop} $pp \ra \tilde{t}_1 \tilde{t}_1 h$ has a large cross 
section and might be a very useful channel for $h$ discovery.}. \s

Let me turn now to the case of $\tb \gg1$, where the off--diagonal entry in the
$\tilde{b}$ mass matrix will play a major role. For instance, choosing 
moderate values for the universal trilinear coupling $A\equiv A_t=A_b$ and the 
common soft--SUSY breaking scalar mass, a large value of the parameter $\mu$ 
[which is then multiplied by $\tb$] will make the off--diagonal entry very 
large, leading to a sizeable splitting between the two sbottom masses with 
$m_{\tilde{b}_1}$ possibly rather small, and a large $ g_{h \tilde{b}_1 
\tilde{b}_1 }$ coupling which could generate large $\tilde{b}_1$ loop 
contributions to the $hgg$ and $h\gamma \gamma$ vertices.
This is shown in Fig.~3, where the effect of the $\tilde{t}$ and $\tilde{b}$ 
loops [for $h \ra \gamma \gamma$
a small contribution is also coming from chargino loops] on the quantities 
R$(h \ra \gamma \gamma)$ and  R$(gg \ra h \ra \gamma \gamma)$ 
is displayed as a function of $\mu$ [with $\mu<0$] for $\tb=50$. 
The values $m_{\tilde{b}_1}=200$ GeV and $M_2=300$ GeV have been chosen.
The thick and thin curves correspond to the two choices $A\equiv \tilde{A}_t  
= \tilde{A}_b=0$ and 0.5 TeV, respectively. \s

The effects of SUSY loops on the $h \ra \gamma \gamma$ decay width is 
relatively small, barely exceeding the level\footnote{For
small values $|\mu| \sim {\cal O}(100)$ GeV, the deviation from unity 
of R$(h \ra \gamma \gamma)$ can be larger but this is mostly due to the 
contribution of the chargino $\chi_1^+$ which, in this case is rather light
\cite{gamma}.} of 10\%.  In turn, the deviations of the R$(h \ra gg)$  
and thus R$(gg \ra 
h \ra \gamma 
\gamma$) observables from unity are substantial for large values of $|\mu|$, 
exceeding a factor of 2 for $|\mu| \sim 800$ GeV. For this $|\mu|$ value 
and above, only the $\tilde{b}$ contribution is sizeable: the $\tilde{t}_1$ 
is either too heavy, or its couplings to the $h$ boson small [this explains why
the two curves for $A=0$ and 0.5 TeV are almost the same]. For lower $\mu$
values, the difference between the two curves is due to the $\tilde{t}_1$ 
contribution.  Thus the effect of sbottom loops on the  observable 
R$(gg \ra \gamma \gamma$) can be sizeable for large values of $|\mu|$.
For extreme values, $|\mu| \simeq 1.2$ TeV [for larger values of $|\mu|$ the 
$h$ boson mass becomes smaller than $90$ GeV], the $gg \ra \gamma \gamma$ 
cross section in the MSSM can be suppressed compared to the SM case by a 
factor of 5. Of course, if the $\tilde{b}_1$ mass is increased (reduced) the 
effect becomes less (more) striking. \s

Finally, a remark on the situation where the decoupling limit is not yet 
reached is in order. In this case the $hWW$ and $htt$ couplings are smaller 
than in the SM,
and both the $gg \ra h$ cross section and $h \ra \gamma \gamma$ widths are 
suppressed compared to the SM case, even in the absence of the squark loops. 
Including light $\tilde{t}$ squark contributions will further decrease
the amplitudes in the case of large $\tilde{A}_t$ as shown in Fig.~4
for $\tb=2.5$ and $M_A=200$ GeV. For large values of $\tb$, the $hgg$ 
amplitude can be enhanced by the $b$--loop contribution, but the $h \ra \gamma 
\gamma$ branching ratio is strongly suppressed due to the absence of the 
$W$--loop and the increase of the total decay 
width $\propto m_b^2 \tan^2\beta$. 
In the case of the heavy CP--even Higgs boson $H$, squark loop contributions
to the cross section $gg \ra H$ can be even larger since because of the larger 
value of $M_H$, more room will be left for the $\tilde{t}$ and $\tilde{b}$ 
squarks before they decouple form the $Hgg$ amplitude. 
In addition, for $M_H$ values above the squark pair threshold, the
decays $H \ra \tilde{t}_1 \tilde{t}_1$ or $H \ra \tilde{b}_1 \tilde{b}_1$
will be kinematically allowed and could have large branching ratios, 
therefore suppressing the other decay modes including the $H \ra \gamma 
\gamma$ channel. For the pseudoscalar Higgs boson, however, squark loops 
will not have
drastic effects on the production cross section $\sigma (gg \ra A)$:
because of CP--invariance, the $A$ boson couples only to $\tilde{t}_1
\tilde{t}_2$ or  $\tilde{b}_1 \tilde{b}_2$ pairs while 
the gluon coupling to different squarks is absent; the $Agg$
amplitude, therefore, cannot be built at lowest order by scalar quark 
loops. 

\subsection*{4. Conclusions}

I discussed the effects of $\tilde{t}$ and $\tilde{b}$ squarks on 
the main production mechanism of the lightest neutral SUSY Higgs boson $h$ at the LHC, 
$gg \ra h$, and on the important decay 
channel $h \ra \gamma \gamma$ in the context of the MSSM. If the off--diagonal
entries in the $\tilde{t}$ and $\tilde{b}$ mass matrices are large, the
eigenstates $\tilde{t}_1$ and $\tilde{b}_1$ can be rather light and at the 
same time their 
couplings to the $h$ boson strongly enhanced. The cross section times 
branching ratio $\sigma( gg \ra h) \times {\rm BR}(h \ra \gamma \gamma)$ 
can be then much smaller than in the SM, even in the decoupling regime, 
$M_A \gg M_Z$, where the $h$--boson has SM--like couplings to fermions and 
gauge bosons. Far from this decoupling limit, the cross section times branching
ratio is further reduced in general due to the additional 
suppression of the $ht\bar{t}$ and $hWW$ couplings. 

\bigskip

\noindent {\bf Acknoweldgements}: \\
I thank Daniel Denegri and Francois Richard for stimulating and fruitful 
discussions.

\begin{figure}[htb]
\hspace*{-1.5cm}
\vspace*{-.5cm}
\mbox{
\psfig{figure=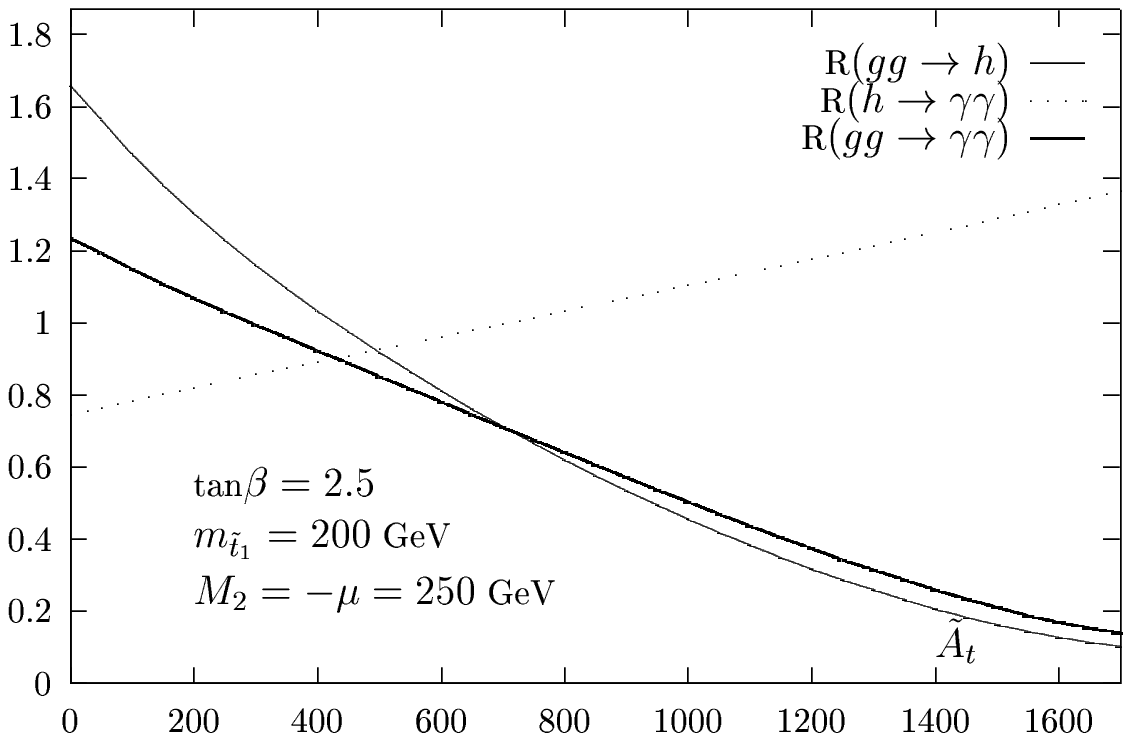,width=18cm}}
\vspace*{-16cm}
\caption[]{SUSY loop effects on R$(h \ra \gamma \gamma)$, 
R$(gg \ra h)$ and their product R$(gg \ra \gamma \gamma)$ as a function of 
$\tilde{A}_{t}$ for $\tb=2.5$ and $m_{\tilde{t}_1}=200$ GeV; $M_2=-\mu=250$  
GeV.}
\end{figure}

\begin{figure}[htb]
\hspace*{-1.5cm}
\vspace*{-.5cm}
\mbox{
\psfig{figure=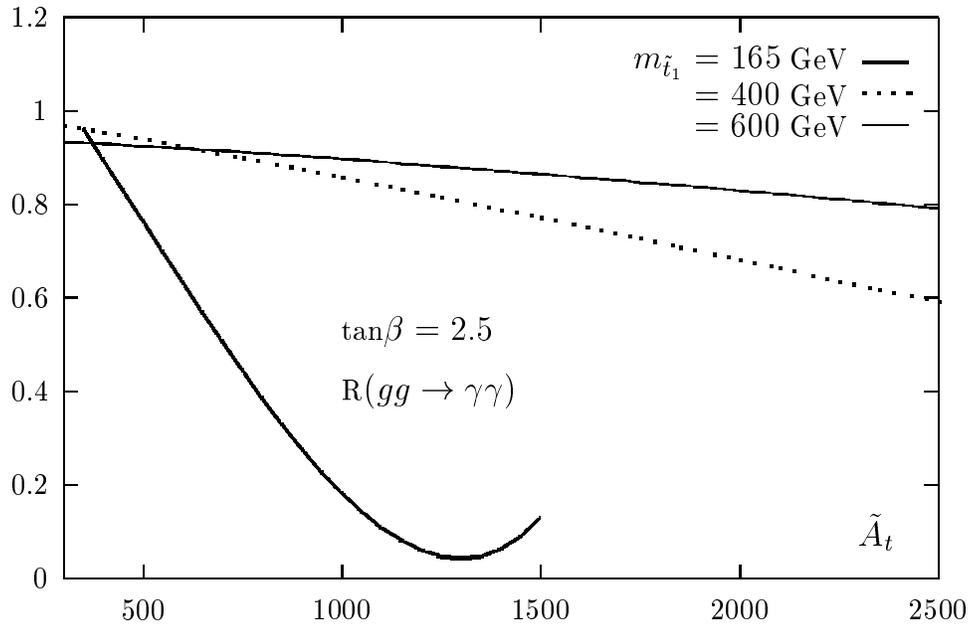,width=18cm}}
\vspace*{-16cm}
\caption[]{SUSY loop effects on R$(gg \ra \gamma \gamma)$ as a 
function of $\tilde{A}_{t}$ for $\tb=2.5$ and $m_{\tilde{t}_1}=165,400$
and 600 GeV.}
\end{figure}

\begin{figure}[htb]
\hspace*{-1.5cm}
\vspace*{-.5cm}
\mbox{
\psfig{figure=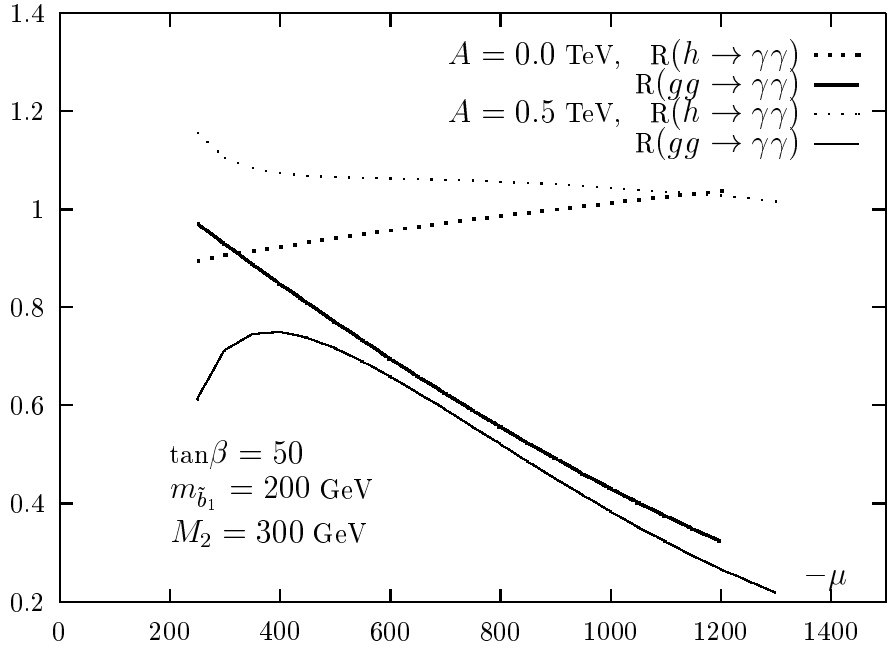,width=17cm}}
\vspace*{-14cm}
\caption[]{SUSY loop effects on R$(h \ra \gamma \gamma)$ and  
R$(gg \ra h \ra \gamma \gamma)$ as a function of $-\mu$ for $\tb=50$ and 
$m_{\tilde{b}_1}=200$ GeV and $A\equiv \tilde{A}_t 
= \tilde{A}_b=0 (0.5)$ TeV for the thick (thin) curves.}
\end{figure}

\begin{figure}[htb]
\hspace*{-1.5cm}
\vspace*{-.5cm}
\mbox{
\psfig{figure=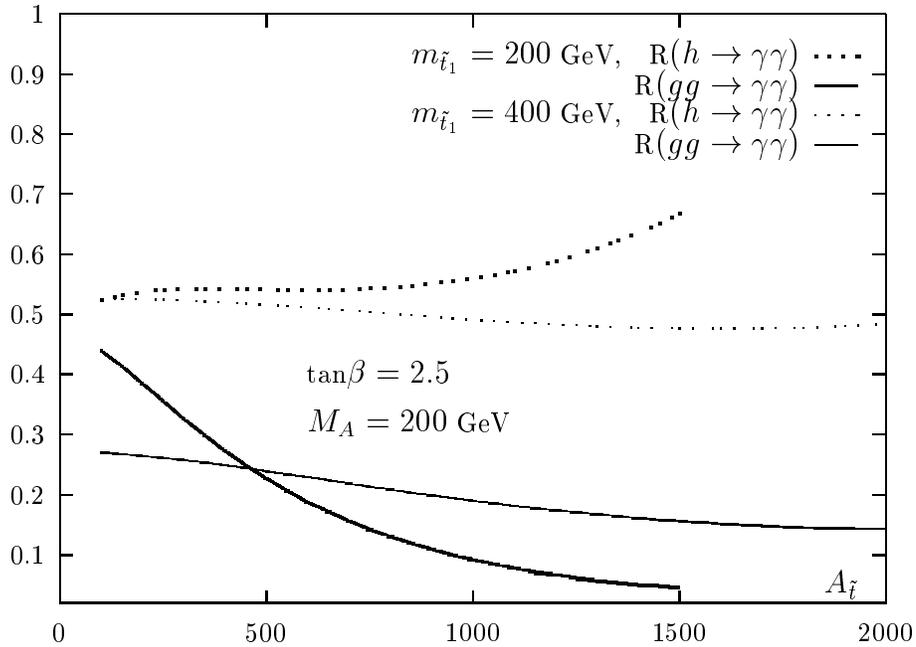,width=17cm}}
\vspace*{-14cm}
\caption[]{SUSY loop effects on R$(h \ra \gamma \gamma)$ and  
R$(gg \ra h \ra \gamma \gamma)$ as a function of $\tilde{A}_t$ for $\tb=2.5,
M_A=200$ GeV and two values  $m_{\tilde{t}_1}=200$ and 400 GeV.}
\end{figure}

\end{document}